\documentclass[12pt]{iopart}
\usepackage{graphicx}
\def\be{\begin{equation}}
\def\ee{\end{equation}}
\def\bea{\begin{eqnarray}}
\def\eea{\end{eqnarray}}

\def\c{\cite}

\def\ov{\over}

\def\rs2r{{r_{s}\over 2r}}
\def\l2r2{{l^{2}\over r^{2}}}

\def\a2{{l^{2}\over a^{2}}}
\def\b2{{l^{2}\over b^{2}}}

\begin{document} 

\jl{6}

\title{Regge Calculus in Teleparallel Gravity}

\author{J G Pereira and T Vargas}
\address{Instituto de F\'{\i}sica Te\'orica\\ 
Universidade Estadual Paulista \\
Rua Pamplona 145 \\
01405-900\, S\~ao Paulo SP \\ 
Brazil}

\begin{abstract}

In the context of the teleparallel equivalent of general relativity, the Weitzenb\"ock 
manifold is considered as the limit of a suitable sequence of discrete lattices composed
of an increasing number of smaller an smaller simplices, where the interior of each simplex
(Delaunay lattice) is assumed to be flat. The link lengths $l$ between any pair of vertices
serve as independent variables, so that torsion turns out to be localized in the two
dimensional hypersurfaces (dislocation triangle, or hinge) of the lattice. Assuming that a 
vector undergoes a dislocation in relation to its initial position as it is parallel
transported along the perimeter of the dual lattice (Voronoi polygon), we obtain the discrete
analogue of the teleparallel action, as well as the corresponding simplicial vacuum field
equations.
 
\end{abstract}

\pacs{04.20.-q; 04.50.+h; 04.60.Nc}

\maketitle

\section{Introduction}

Regge calculus is a useful tool for numerical calculation in curved spacetimes, and has been
applied to a variety of problems in cosmology, as well as in classical and quantum
gravity~\c{wi}. In the well known paper by Regge~\c{re}, the usual continuous space-time
manifold of general  relativity is viewed as the limit of a suitable sequence of discrete
lattices composed of an increasing number of smaller an smaller simplices. In this approach,
the interior of each simplex is assumed to be flat, and the link lengths $l$ between any pair
of its vertices serve as independent variables so that the scalar curvature is defined as a
distribution with support on the bones of the lattice.

Few attempts have been made to include concepts of non-riemannian geometry into the 
Regge calculus~\cite{c1,c2,c3,c4,c5,c6}. In some of these attempts~\c{c2,c5}, the Regge calculus
was formulated as a lattice gauge theory for the Poincar\'e group, and the possibility of
including torsion as closure failures of the building blocks of the simplicial manifold was
pointed out. Another of such attempt was an application to the Einstein-Cartan theory~\c{c6},
where the square of the torsion tensor was defined as a distribution, and the corresponding
simplicial field equations were obtained. The notion of torsion singularities appearing as a
conical defect (dislocation) has also been discussed in the literature~\c{t}. Its application to
the theory of solid state and  crystal defects has been known for a long time~\c{k,b,kr,ga,h},
and the connection to the low dimensional gravity has already been pointed out~\c{ho,ka,kl}.    
       
An alternative approach to gravitation is the so called teleparallel gravity~\c{M}, which
is gauge theory based on the Weitzenb\"ock geometry~\c{we}. In this theory, gravitation is
attributed to torsion~\cite{xa}, which plays the role of a force~\cite{pe}, and the
curvature tensor vanishes identically. As is well known, at least in the absence of spinor
fields, teleparallel gravity is equivalent to general relativity. In this paper, relying upon
this equivalence, we will construct the discrete counterpart of the four-dimensional
teleparallel action, and will obtain the corresponding field equations in vacuum. In a vacuum
simplicial Weitzenb\"ock geometry, the torsion tensor is localized in the two-dimensional
dislocation triangles, called hinges. Torsion can then be detected by measuring the dislocation
in relation to the initial position of a vector, as a result of the parallel transport along a
small loop encircling the dislocation triangle (hinge), where torsion is concentrated. When
torsion is present, it is detected parallel to this hinge, and the dislocation is measured by
the Burgers vector $b_d$. The metric properties of this simplicial manifold are given by the
lengths ${l_p}$ of the edges. This means that the variation of the simplicial action is to be
made with respect to ${l_p}$. The resulting simplicial field equations can be considered
as the teleparallel equivalent of the simplicial Einstein's equation of general relativity,
first obtained by Regge~\cite{re}.

It should be remarked that the Regge calculus has already been applied to the
Einstein-Cartan theory~\c{c6}, where the fermions act as a source of torsion. In this case, the
Burgers vector $(b_d)$ couples algebraically to the matter term, and so the variation of the
action is to be made with respect to both the edge lengths and the Burgers vectors. Furthermore,
as in the absence of matter no dislocations can be present on the lattice, the Burgers vector
will be zero, and the simplicial field equations will reduce to the usual Regge's equations of
general relativity. As we are going to see, in the teleparallel approach, even in the presence
of matter, the Burgers vector does not couple to matter. Furthermore, in the absence of matter,
the Burgers vector does not vanish, a property which is in accordance with the fact that, in
teleparallel gravity, torsion is a propagating field. We will proceed according to the following
scheme. In section~2, we review the main features of teleparallel gravity. In section~3, we
obtain the simplicial torsion, the discrete action, and the simplicial field equation.
Discussions and conclusions are presented in section~4.

\section{Teleparallel Equivalent of General Relativity} 

It is well known that curvature, according to general relativity, is used to geometrize 
the gravitational interaction. On the other hand, teleparallelism attributes gravitation to
torsion, but in this case torsion accounts for gravitation not by geometrizing the interaction,
but by acting as a force~\c{pe}. This means that in the teleparallel equivalent of general
relativity, instead  of geodesics, there are force equations quite analogous to the Lorentz
force equation of electrodynamics.

A nontrivial tetrad field induces on spacetime a teleparallel structure which is directly
related to the presence of the gravitational field. In this case, tensor and local Lorentz
indices\footnote{The greek alphabet ($\mu$, $\nu$, $\rho$,~$\cdots=0,1,2,3$) will be used to
denote tensor indices, that is, indices related to spacetime. The latin alphabet ($a$, $b$,
$c$,~$\cdots=0,1,2,3$) will be used to denote local Lorentz (or tangent space) indices, whose
metric tensor is chosen to be $\eta_{ab} = \mbox{diag} (+1, -1, -1, -1)$. Furthermore, we will
use units in which $\hbar = c =1$.} can be changed into each other with the use of a tetrad field
$h^{a}{}_{\mu}$. A nontrivial tetrad field can be used to define the linear Weitzenb\"ock
connection 
\be
\Gamma^{\sigma}{}_{\mu \nu} = h_a{}^\sigma \partial_\nu h^a{}_\mu,
\label{car}
\ee
a connection presenting torsion, but no curvature. It parallel transports the tetrad itself: 
\be
{\nabla}_\nu \; h^{a}{}_{\mu}
\equiv \partial_\nu h^{a}{}_{\mu} - \Gamma^{\rho}{}_{\mu \nu}
\, h^{a}{}_{\rho} = 0. 
\label{weitz}
\ee 
The Weitzenb\"ock connection satisfies the relation 
\be
{\Gamma}^{\sigma}{}_{\mu \nu} = {\stackrel{\circ}{\Gamma}}{}^{\sigma}{}_{\mu
\nu} + {K}^{\sigma}{}_{\mu \nu},
\label{rel} 
\ee
where
\be
{\stackrel{\circ}{\Gamma}}{}^{\sigma}{}_{\mu \nu} = \frac{1}{2}
g^{\sigma \rho} \left[ \partial_{\mu} g_{\rho \nu} + \partial_{\nu}
g_{\rho \mu} - \partial_{\rho} g_{\mu \nu} \right]
\label{lci}
\ee
is the Levi--Civita connection of the metric 
\be
g_{\mu \nu} = \eta_{a b} \; h^a{}_\mu \; h^b{}_\nu,
\label{gmn}
\ee
and 
\be {K}^{\sigma}{}_{\mu \nu} = \frac{1}{2}
\left[T_{\mu}{}^{\sigma}{}_{\nu} +T_{\nu}{}^{\sigma}{}_{\mu}-T^{\sigma}{}_{\mu \nu}\right]
\label{conto}
\ee 
is the contorsion tensor, with 
\be
T^\sigma{}_{\mu \nu} =
\Gamma^{\sigma}{}_{\nu \mu} - \Gamma^ {\sigma}{}_{\mu \nu} \;  \label{tor} 
\ee
the torsion of the Weitzenb\"ock connection~\c{ald}.

The teleparallel gravitational lagrangian is
\be
L_{G} = \frac{h}{16 \pi G}\left[{1\ov 4} T^{\rho}{}_{\mu \nu}T_{\rho}{}^{\mu \nu} + 
{1\ov 2} T^{\rho}{}_{\mu \nu}T^{\nu \mu}{}_{\rho} - T_{\rho \mu}{}^{\rho}T^{\nu \mu}{}_{\nu}
\right],
\label{la}
\ee
where $h = \det(h^{a}{}_{\mu})$. By considering now the functional variation of
$L_G$ in relation to $h^{a}{}_{\mu}$, we obtain the teleparallel gravitational field
equation~\c{an}
\be
\partial_{\sigma}\left(hS_{\lambda}{}^{\sigma \tau}\right)-4 \pi G(ht_{\lambda}{}^{\tau})=0,
\ee
where 
\be
t_{\lambda}{}^{\tau}=\frac{1}{4\pi G} \, \Gamma^{\mu}{}_{\nu \sigma} \,
S_{\mu}{}^{\sigma \tau} + 
\delta_{\lambda}{}^{\tau}L_G,
\ee
is the canonical energy-momentum pseudo tensor of the gravitational field, and $S^{\rho \mu
\nu}$ is the tensor
\be
S^{\rho \mu \nu}=\frac{1}{2}\left[ K^{\mu \nu \rho}- g^{\rho \nu}T^{\theta \mu}{}_{\theta} +
g^{\rho \mu}T^{\theta \nu}{}_{\theta}\right].
\ee
 
\section{Discrete torsion and simplicial field equation}

We are going to obtain now the simplicial teleparallel action, as well as the corresponding
field equation. In a heuristic way, and similarly to the Regge construction of the simplicial
manifold of general relativity, we assume that the usual continuous spacetime manifold of
teleparallel gravity can be viewed as the limit of a suitable sequence of discrete lattices
composed of an increasing number of smaller an smaller simplices. In other words, the
Weitzenb\"ock manifold, which is the stage set of teleparallel gravity, is approximated by a
four-dimensional polyhedra $M^4$. In this approach, the interior of each simplex is assumed to
be flat, and this flat four-simplices are joined together at the tetrahedral faces of their
boundaries. The torsion turns out to be localized in the two-dimensional dislocation triangles
(hinges) of the lattice, and the link lengths $l$  between any pair of vertices serve  as
independent variables. By varying the action with  respect to this edge lengths, one obtains the
simplicial analogues of the teleparallel field equations.

Let us then proceed to this construction. To begin with, let us take a bundle of parallel
dislocations (hinges) in $M^3$. We make the assumption that the torsion induced by the
dislocations is small, so that we may regard $M^3$ as approximately euclidian. Let ${\bf U}$ be
a unity vector parallel to the dislocations. We test for the presence of torsion by carrying a
vector ${\bf A}$ around a small loop of area vector ${\bf S}= S{\bf n}$, with $S$ the area and
${\bf n}$ a unity vector normal to the surface. At the end of the test, if torsion is
nonvanishing, ${\bf A}$ is found to have translated from the original position, along ${\bf U}$,
by the length ${\bf B} = N {\bf b}$, where $N$ is the number of dislocations entangled by the
loop, and ${\bf b}$ is the Burgers vector, which is a vector that gives both the length and
direction of the closure failure for every dislocation. In $M^4$, the flux of dislocation lines
through the loop of area $S^{\alpha \beta}$ is
\[
\Phi = \rho \left({\bf U}{\bf S} \right) = {1\over2} \, \rho_{\alpha \beta} \, S^{\alpha \beta},
\]
where $\rho$ is the density of dislocation passing through the loop, and $\rho_{\alpha \beta} = 
\rho \, U_{\alpha \beta}$, with $U^{\alpha \beta}$ a unity antisymmetric tensor satisfying
$U_{\alpha \beta}U^{\alpha \beta} = 2$. This means that
we  can endow the polyhedra more densely with hinges in a region of high torsion than in
region of low torsion. The closure failure is then found to be
\be
B_{\mu} = {1\over2} \, \rho_{\alpha \beta} \, S^{\alpha \beta}\,b_{\mu}.       
\ee
However, we know from differential geometry that, in the presence of torsion, 
infinitesimal parallelograms in spacetime do not close, the closure failure being equal to
\be
B_{\mu}=T_{\mu \nu \sigma} \, {S}^{\nu \sigma}.
\label{bts}
\ee
By comparing the last two equations we see that
\be
T_{\mu \alpha \beta} = \frac{1}{2} \, \rho_{\alpha \beta} \,b_{\mu} \equiv
\frac{1}{2} \, \rho \, U_{\alpha \beta} \,b_{\mu}.
\label{tordef}
\ee

On the other hand, it has already been shown~\c{t} that the torsion singularity takes the form
of a conical  singularity. Consequently, the dislocation from the original position that occurs
when a vector is parallel transported around a small loop encircling a given bone is independent
of the area of the loop. Furthermore, the dislocation has two characteristics: The length of the
dislocation line in three-dimensions, and the area of the triangle in four-dimensions.
Therefore, there is a natural unique volume associated with each  dislocation. To define this
volume, there is a well-known procedure in which a  dual lattice is constructed for any given
lattice~\c{ch,mi}. This involves constructing polyhedral cells around each vertex, known in
the literature as Voronoi polygon, in such a way that the polygon around each particular
vertex contains all points which are nearer to that vertex than to any other vertex. The
boundary of the Voronoi polygon is always perpendicular to the edges emanating from the
vertex, and each corner of the Voronoi polygon lies at the circuncentre of any of the
simplices of the Delaunay geometry, which shares the dislocation (bone) (see Fig.~1).

%%%%%%%%%%%%%%%%%%%
\begin{figure}
\begin{center}
\scalebox{0.35}[0.35]{\includegraphics{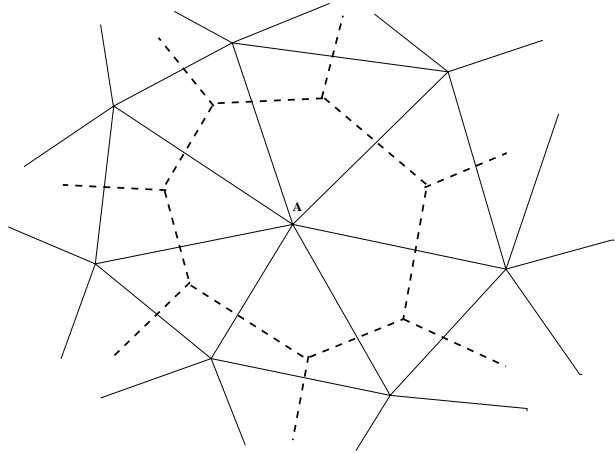}}
\end{center}
\caption{The two-dimensional Voronoi polygon (dashed line) for a particular vertex $A$. Each 
corner of the Voronoi polygon lies at the circuncentre of any of the triangles of the Delaunay 
geometry (solid line).}
\label{1}
\end{figure}
%%%%%%%%%%%%%%%%%%%

By construction, the Voronoi polygon is orthogonal to the bone. If we parallel transport a
vector around the perimeter of a Voronoi polygon of area ${\Sigma^{\ast}_{d}}$, it will
traverse the flat geometry of the interior of each one of the simplices sharing the bone, and
will return dislocated from its original position in a plane parallel to the bone by a length
$b_{\mu}$. According to this construction, and relying on the definition (\ref{tordef}), the
torsion due to each dislocation can be expressed by (see Fig.~2):
\be
\mbox{(Torsion)} = \frac{\mbox{(Distance the vector is translated)}}{\mbox{(Area
circumnavigated)}}.
\ee
This definition is equivalent to the definition of Gauss curvature:
\be
\mbox{(Gauss curvature)} = \frac{\mbox{(Angle the vector is rotated)}}{\mbox{(Area
circumnavigated)}}.
\ee
Therefore, analogously to the Riemann scalar, which is proportional to the Gauss curvature, and
whose proportionality constant depends on the dimension $D$ of the lattice geometry~\c{mi},
we define the simplicial torsion due to each dislocation by
\be
T_{(d)\mu \nu \rho} = \sqrt{D(D-1)} \ \; \frac{b_{(d)\mu}U_{(d)\nu \rho}}{{\Sigma^{\ast}_{d}}}
\equiv \sqrt{12} \ \; \frac{b_{(d)\mu}U_{(d)\nu \rho}}{{\Sigma^{\ast}_{d}}}.
\label{T}
\ee
The reason for the square root is that, as is well known from teleparallel gravity, the Riemann
curvature tensor is proportional to a combination of squared torsion tensors.

As we have said, the vector returns translated from its original position in a plane parallel
to the hinge by a length $b^{\mu}$. Let us then analyze the translational group acting on it.
As we know, the interior of each block is flat (Minkowski space), so the infinitesimal
translation in these blocks is given by
\be
T(\delta b) = I - i \; \delta b^{a} \, P_{a}, 
\ee    
where $I$ is the unity matrix, $\delta b^{a}$ are the components of an arbitrarily small
four-dimensional Burgers (displacement) vector, and $P_{a} = i \partial_{a}$ are the translation
generators. In the presence of dislocations, and using the tetrad $h^{a}{}_{\mu}$, this
infinitesimal translation parallel to the hinge becomes
\be
T(\delta b) = I - i \; \delta b^{\mu} \, h^{a}{}_{\mu} \, P_{a}, 
\ee 
so that a finite translation will be represented by the group element
\be
T(b) = \exp \left[ -i \, b^{\mu} \, h^{a}{}_{\mu} \, P_{a} \right]. 
\ee
On the other hand, the contour integral of the Burgers vector --- which measures how much the
infinitesimal closed contour $\Gamma$ spanning a surface element
$d \Sigma^{\ast \alpha \beta}$ fails to close in the presence of hinge --- by using
Eq.~(\ref{bts}), is seen to be~\c{kle}
\be
b^{\mu}= \oint\limits_{\Gamma} T^{\mu}{}_{\alpha \beta} \, d \Sigma^{\ast \alpha \beta}.
\ee
Therefore, the group element of translations due to torsion turns out to be 
\be
T(b) = \exp \left[-i\oint\limits_{\Gamma} P_{a} \, T^{a}{}_{\alpha \beta} \,
d \Sigma^{\ast \alpha \beta} \right]. 
\ee
%%%%%%%%%%%%%%%%%%%
\begin{figure}
\begin{center}
\scalebox{0.35}[0.35]{\includegraphics{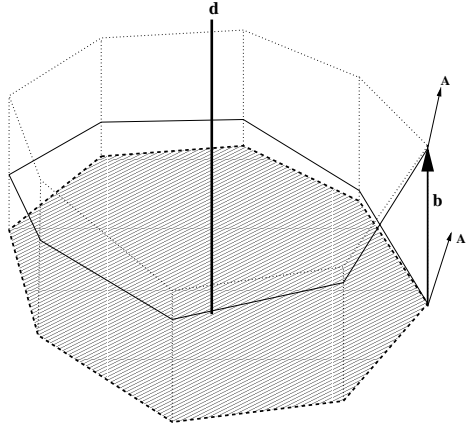}}
\end{center}
\caption{The parallel transport of a vector $A$ along the perimeter of a Voronoi polygon (dashed 
line) around the dislocation line in three-dimension. The vector returns translated in a plane
parallel to the dislocation $d$ by the length $b$.}
\label{2}
\end{figure}
%%%%%%%%%%%%%%%%%%%
      
The four-volume $\Omega_d$ associated with each dislocation, as described above, is 
proportional to the product of $\Sigma_d$, the two-dimensional volume of the dislocation, and
$\Sigma^{\ast}_d$, the area of the Voronoi polygon~\c{mi}:
\be
\Omega_d \equiv \frac{2}{D(D-1)} \; \Sigma_d \, {\Sigma^{\ast}_d} =
\frac{1}{6} \; \Sigma_d \, {\Sigma^{\ast}_d}.
\ee
The invariant volume element $h \, d^{4}x$, therefore, is represented by $\Omega_d$, and we
have the following relation,
\be
\int h \, d^{4}x \Longrightarrow \sum_{\rm dis}\Omega_d = \frac{1}{6} \;
\sum_{\rm dis} \Sigma_d \, {\Sigma^{\ast}_d},
\ee
where the sum is made over all dislocations. We are ready then for constructing the simplicial
action. Let us take the lagrangian (\ref{la}) of teleparallel gravity, whose terms are
proportional to the square of the torsion tensor, and substitute torsion as given by
Eq.~(\ref{T}). For the first term, we obtain
\be
T_{(d)}^{\mu \nu \rho} \, T_{(d)\mu \nu \rho} =
24 \left(\frac{1}{{\Sigma^{\ast}_d}} \right)^{2}b^{\mu}_{(d)} \, b_{(d) \mu}.
\ee
Writing the other two terms in a similar way, the simplicial teleparallel action will be 
\be
S = \frac{1}{16 \pi G}\sum_{\rm dis} \left(\frac{b^2_d}{{\Sigma^{\ast}_d}}\right)
\Sigma_d,
\label{tpa}
\ee
where $b^{2}_{d}$ denotes the projected Burgers vector parallel to the hinge, 
and $\Sigma_d$ is the area of this dislocation triangle, or hinge.

On the other hand, the simplicial Einstein-Hilbert action of general relativity is~\cite{re}
\be
S = \frac{1}{8 \pi G} \sum_{\rm hinge} \; \varepsilon_h \; \Sigma_h,
\label{rga}
\ee
where $\varepsilon_h$ is the deficit angle associated to each hinge, which is directly related
to the curvature of spacetime. Now, up to a surface term, the Einstein-Hilbert action of
general relativity is known to be equivalent to the teleparallel action (\ref{la}). Assuming
that the corresponding simplicial versions of the same actions are also equivalent, we can
obtain a relation between the {\it angle} $\varepsilon_h$, through which a vector is {\it
rotated} in relation to its initial direction  when parallel transported around a small loop,
and the {\it distance} $b_d$ through which a vector is {\it translated} from its original
position when parallel transported around a small loop. By comparing Eqs.~(\ref{tpa}) and
(\ref{rga}), this relation is found to be
\be
\varepsilon_h = \frac{1}{2}\left(\frac{b^2_d}{{\Sigma^{\ast}_d}}\right).
\label{simeq}
\ee

Let us now carry out the variation of the simplicial action (\ref{tpa}) with respect to the
edge lengths $l_p$. In the usual application of Regge calculus to a curved spacetime, the
typical deficit angle $\varepsilon_h$ related to each hinge depends in a complicated
trigonometric way on the values of numerous edge lengths $l_p$. However, as demonstrated by
Regge~\cite{re}, one can carry out the variation with respect to the edge lengths $l_p$ as if
the $\varepsilon_h$ were constants. Similarly, in the teleparallel simplicial gravity, both the
projected Burgers vector $b^2_d$ and the area of the dual Voronoi polygon $\Sigma^{\ast}_d$
related to each hinge depend also in a complicated trigonometric way on the values of numerous
edge lengths $l_p$. Because of the relation (\ref{simeq}), however, in the same way
$\varepsilon_h$ can be considered as a constant in the variation of the simplicial
Einstein-Hilbert action, one can equivalently carry out the variation of the teleparallel
simplicial action (\ref{tpa}) as if the dimensionless parameters $({b^2_d}/{{\Sigma^{\ast}_d}})$
were constants. This is the teleparallel version of the result demonstrated by Regge. We thus
find
\be
\delta S = \frac{1}{16 \pi G}\sum_{\rm dis} \left(\frac{b^2_d}{{\Sigma^{\ast}_d}} \right)
\delta \Sigma_d.
\ee
The change in the area of the triangle hinge, on the other hand, is given by
\be
\delta \Sigma_d = \frac{1}{2} \sum_{p} \, l_{p} \, \delta l_{p} \, {\rm cot} \theta_{(dp)}, 
\ee
where $\theta_{(dp)}$ is the angle in the triangle hinge $d$ opposite to the edge $l_p$. The
simplicial teleparallel field equation in vacuum then reads 
\be
l_{p} \sum_{d\supset l_{p}} \left(\frac{b^2_d }{{\Sigma^{\ast}_d}}\right){\rm cot} 
\theta_{(dp)} = 0,
\label{tseq}
\ee
where the sum is made over all dislocation triangles $d$ which have the given edge $p$ in
common.

The simplicial equation (\ref{tseq}) is a set of equations for the edge lengths $l_p$, and
consequently there will be one simplicial equation for each edge in the lattice. To solve it,
initial conditions consisting of informations about some of the edge lengths must be supplied.
Then, from inputs expressing conditions on the torsion, that is, on the Burgers vector
associated to each hinge, one can determine the remainder of the edge lengths, and consequently
the geometrical properties of the simplicial Weitzenb\"ock manifold. This is quite similar to
simplicial general relativity, where the inputs are conditions on the curvature, that is, on the
deficit angle associated to each hinge. It should be remarked, however, that despite being as
numerous as the edges, the simplicial equations are not able in general to determine all edge
lengths. The problem is that there are relations between these equations that prevent them from
being independent, and in consequence, they cannot determine all the details of the largely
arbitrary lattice. Therefore, to prepare a real computer program based on Regge calculus, one
has to supply the computer not only with the simplicial equation and initial conditions, but
also with definite algorithms to remove all the arbitrariness related to the lattice being
used~\cite{mtw}.

\section{Final Remarks}

We have considered in this paper the Weitzenb\"ock geometry as the limit of a suitable
sequence of discrete lattices, the interior of each simplex being assumed to be flat, and the
torsion localized in the two-dimensional dislocation triangles (hinges) of the lattice. The
information about the metric properties is codified in the link lengths $l_p$ between any 
pair of vertices. Furthermore, from the fact that a parallel transported vector undergoes a
translation from its original position as it is transported along the perimeter of the Voronoi
polygon, we have found the simplicial torsion. Analyzing the infinitesimal translations in flat
lattices, the translational group  due to torsion was also found. Then, by using the simplicial
torsion, we have constructed the simplicial version of the teleparallel action, and finally, by
varying the simplicial action with respect to the edge lengths, we have found the simplicial
field equations in vacuum.

In the simplicial Einstein-Cartan approach, the fermions act as a source of torsion, and
the Burgers vector couples to the matter term. In the  absence of matter, there is no
dislocation on the lattice, and the Burgers vector is consequently zero. In this case, the
simplicial field equations reduce to the Regge's equation of general relativity. On the other
hand, in our approach, the geometrical properties of the vacuum simplicial Weitzenb\"ock
manifold is replaced by the lengths ${l_p}$ of the edges, which  means that the variation of the
simplicial action is to be made with respect to ${l_p}$ as if both the Burgers vector and the
area of the Voronoi dual polygon were constants. The corresponding simplicial field equations
turns out to be the teleparallel equivalent of the Regge's general relativity equations. It
should be remarked that in this approach, even in the presence of matter, the Burgers vector
does not couple to matter, and the variation is made only with respect to ${l_p}$. Furthermore,
differently from the simplicial Einstein-Cartan approach, in the absence of matter, the Burgers
vector does not need to vanish, a property that reflects the fact that, in teleparallel gravity,
torsion is a propagating field. Due to its gauge structure, teleparallel gravity turns out to be
quite similar to the other known gauge theories, and depending on the problem to be considered,
it can present several formal advantages in relation to general relativity. In this connection,
we hope that these results can be useful for numerical calculations in different contexts like
gravitation, cosmology, as well as quantum gravity.

\ack
The authors would like to thank Yu. N. Obukhov for useful comments. They would like to thank
also FAPESP-Brazil, CNPq-Brazil and CAPES-Brazil for financial support.

\end{document}